\documentclass[prd,twocolumn,superscriptaddress,altaffilletter,amssymb,showpacs,nofootinbib]{revtex4}
\usepackage[dvips]{graphicx}
\usepackage{amsmath}
\usepackage{graphicx}
\usepackage{amsmath,amssymb}        
\usepackage{epsfig}
\usepackage{bm}
\usepackage{amsfonts}
\usepackage{amsmath,amssymb}
\usepackage{dcolumn}
\PassOptionsToPackage{linktocpage}{hyperref}
\usepackage[draft=false]{hyperref}
\usepackage{threeparttable}
\usepackage[draft=false]{hyperref}
\usepackage{threeparttable}
\newcommand{\sfrac}[2]{{\textstyle{#1\over#2}}}
\def\case#1/#2{\textstyle\frac{#1}{#2}}
\newcommand{\ds}{\displaystyle}
\newcommand{\be}{\begin{equation}}
\newcommand{\ee}{\end{equation}}
\newcommand{\ben}{\begin{eqnarray}}
\newcommand{\een}{\end{eqnarray}}
\newcommand{\vphi}{\varphi}

\newtheorem{thm}{Theorem}[section]

\newtheorem{prop}[thm]{Proposition}

\begin{document}

\title{Equilibrium sets in quintom cosmologies: the past asymptotic dynamics}
\author{Genly Leon}\email{genly@uclv.edu.cu}\affiliation{Department of Mathematics, Universidad Central de Las Villas, Santa Clara CP 54830, Cuba}
\author{Rolando Cardenas}\email{rcardenas@uclv.edu.cu}\affiliation{Department of Physics, Universidad Central de Las Villas, Santa Clara CP 54830, Cuba}
\author{Jorge Luis Morales}\email{jmm@uclv.edu.cu}\affiliation{Department of Mathematics, Universidad Central de Las Villas, Santa Clara CP 54830, Cuba}
\date{\today}

\begin{abstract}
In the previous paper \cite{Lazkoz:2006pa} was investigated the phase space of quintom cosmologies for a class of exponential potentials. This study suggests that the past asymptotic dynamics of such a model can be approximated by the dynamics near a hyperbola of critical points. In this paper we obtain a normal form expansion near a fixed point located on this equilibrium set.
We get computationally treatable system up to fourth order. From the structure of the unstable manifold (up to fourth order) we see that that the past asymptotic behavior of this model is given by a massless scalar field cosmology for an open set of orbits (matching the numerical results given in \cite{Lazkoz:2006pa}). We complement the results discussed there by including the analysis at infinity. Although there exists unbounded orbits towards the past, by examining the orbits at infinity, we get that the sources satisfy the evolution rates $\dot\phi^2/V\sim\frac{2m^2}{n^2-m^2},\, \dot\phi/\dot\varphi\sim-m/n,$ with $H/\dot\phi$ approaching zero.
\end{abstract}

\maketitle

\section{Introduction}

Since 1998, the observational evidence coming from type I-a
Supernovae (SNIa), Large Scale Structure (LSS) formation, and the
Cosmic Microwave Background (CMBR) Radiation \cite{obs} for an
expanding Universe is more and more convincing. The acceleration
of the expansion seems to be fuelled by a yet unknown kind of
matter named dark energy (see the reviews \cite{dark}). The most
promising particle candidates for dark energy are spin zero
bosons, which in the mathematical formalism of Quantum Field
Theory are represented by scalar fields. Many models of dark
energy use a single scalar field and two very popular options are
the so called phantom field and the quintessence field.

Beyond the theories with a single scalar field, models with two
fields (quintessence and phantom) have settled out explicitly and
named quintom models
\cite{quintom,Wei:2005fq,Wei:2005nw,Guo:2004fq,Zhang:2005eg,Lazkoz:2006pa,stringinspired,Cai:2008gk,Saridakis:2009uu,
Lazkoz2007,arbitrary}. The quintom paradigm is a hybrid construction of a quintessence
component, usually modelled by a real scalar field that is
minimally coupled to gravity, and a phantom field: a real scalar
field --minimally coupled to gravity-- with negative kinetic
energy. Let us define the equation of state parameter of any
cosmological fluid as $w\equiv\text{pressure}/\text{density}$. The
simplest model of dark energy (vacuum energy or cosmological
constant) is assumed to have $w =-1$. A key feature of
quintom-like behavior is the crossing of the so called phantom
divide, in which the equation of state parameter crosses through
the value $w=-1.$ 
Quintom behavior (i.e., the $w=-1$ crossing) has been investigated
in the context of h-essence cosmologies
\cite{Wei:2005fq,Wei:2005nw}; in the context of holographic dark
energy \cite{holographic};
inspired by string theory
\cite{stringinspired}; derived
from spinor matter \cite{Cai:2008gk}; for arbitrary potentials
\cite{Lazkoz2007,arbitrary}; using
isomorphic models consisting of three coupled oscillators, one of
which carries negative kinetic energy (particularly for
investigating the dynamical behavior of massless
quintom)\cite{setare1}. The crossing of the phantom divide is also
possible in the context of scalar tensor theories
\cite{Elizalde2004,Apostolopoulos:2006si,Bamba:2008xa,Bamba:2008hq,Setare:2008mb}
as well as in modified theories of gravity \cite{Nojiri:2006ri}.

The cosmological evolution of quintom model with exponential
potential has been examined, from the dynamical systems viewpoint,
in \cite{Guo:2004fq} and \cite{Zhang:2005eg,Lazkoz:2006pa}. The
difference between \cite{Guo:2004fq} and
\cite{Zhang:2005eg,Lazkoz:2006pa} is that in the second case the
potential considers the interaction between the conventional
scalar field and the phantom field. In \cite{Lazkoz:2006pa} the case in which the interaction term
dominates against the mixed terms of the potential, was studied.
It was proven there, as a difference with the results in \cite{Zhang:2005eg}, that the existence of scaling attractors (in which the
energy density of the quintom field and the energy density of DM
are proportional) excludes the existence of phantom attractors. Some of this results were extended in
\cite{Lazkoz2007}, for arbitrary potentials. 

In \cite{Lazkoz:2006pa} were investigated curves  of
(non-hyperbolic) critical points $C_\pm,$ and other types of
hyperbolic critical points from the dynamical systems perspective.
We will focus here on the curves $C_\pm.$ These curves are
associated with the past dynamics of quintom models
\cite{Lazkoz:2006pa}. This fact was supported by numerical
integrations. Although can be argue that these curves are
non-physical because the value $w=1,$ we believe that the
investigation of the past dynamics of such system is important for
completeness. This involve the analysis of sets of non-hyperbolic
critical points.

When a vector field ${\bf x}'={\bf X}({\bf x})$ in $\mathbb{R}^n,$
admits non-isolated critical points, for instance a curve $C$ of
critical points, then the matrix of derivatives ${\bf D X(x)},$
when evaluated on each point in the equilibrium set $C,$ has
necessarily one zero eigenvalue. Hence, each point in $C$ is
non-hyperbolic. Then, linealization technique fails to be applied.
Although the equilibrium points
in $C$ are non-hyperbolic, one is able to apply the Invariant
Manifold Theorem \cite{reza,arrowsmith,wiggins} to guaranteed
conditions for the existence of  stable and unstable manifolds.
Suppose that each point in the equilibrium set $C,$ assumed to be
a curve, has an unstable manifold of dimension $n_u,$ the union of
these manifolds forms an $(n_u+1)$-dimensional set whose orbits
approach a point of $C$ as $t\rightarrow -\infty.$ If $n_u=n-1$ we
say that $C$ is a local source \cite{reza}. If the equilibrium set
has only one zero eigenvalue and the others with non-zero real
parts, then the equilibrium set is called normally hyperbolic
\cite{aulbach}. In such a case one is able to obtain useful
information about the stability of the critical set by examining
the signs of the other, non-null, eigenvalues, as we will see
later. However, if one is interested in the construction of center
manifolds, or at least approximated ones, if conditions for their
existence are fulfilled (for references on center manifolds see
\cite{arrowsmith}, chapter 4; \cite{carr}, \cite{wiggins}(b),
chapter 18), we require to use other tools, e.g., normal forms. 

For the construction of (at least, approximated) invariant manifolds
for critical points, we can transform a dynamical system defined in
the neighborhood of a critical point to a non-linear dynamical
system in its simpler form. The theory of Normal Forms (NF) can be
applied in order to do so, and in most occasions, terms of order
higher than two in the Taylor expansion are required. For the
construction of normal forms for vector fields in $\mathbb{R}^n$ we
will follow the approach in \cite{arrowsmith}, chapter 2. NF
calculations consist of two stages: first, the construction of the
normal forms in which case the nonlinear terms take their simplest
form and second, the determination of the topological types of the
fixed points from the normal forms. Then by using techniques
described in \cite{wiggins}(b), we will able to construct
approximated invariant manifolds up to the desired order in the
vector norm.

In this paper we try and investigate Normal Forms expansions of the system in
\cite{Lazkoz:2006pa} near an hyperbolae of critical points. Our
main purpose is to devise approximated unstable and center manifolds. We develop the normal expansion
(with undetermined coefficients) up to an arbitrary order. The
system resulting by truncating error terms admits  a solution
given in quadratures, which in general cannot be solved
explicitly. By integrating the resulting system, one is able to
construct (theoretical) approximated unstable and center
manifolds. The problem to determine all the coefficient values, up
to a prescribed order, if very difficult. Our computing power
allow us to determine the coefficients values up to fourth order.

The paper has been organized as follows: in section \ref{frame} we
make a general description of the model under study. In section
\ref{sectionIII} we recall and improve some results from reference
\cite{Lazkoz:2006pa}. To make the paper self-contained we
explicitly state in section \ref{NFTheory} the main techniques of
the NF expansions by following the approach in \cite{arrowsmith}
(see also \cite{wiggins}(b), chapter 19). We apply these
techniques to the quintom scalar field solution in section
\ref{normal} (which we argue are associated to the past asymptotic
dynamics). In section \ref{infinite} we offer the analysis at
infinity obtaining sufficient conditions for the existence of
local sources at this regime. In section \ref{conclusions} we
offer our main conclusions.

\section{The quintom model}\label{frame}

In the following discussion on the quintom phase space analysis we
restrict ourselves to the two-field quintom model, with a
Lagrangian:
\begin{equation}\label{expquintomlag}
    \mathcal{L}=\frac{1}{2}\partial_{\mu}\phi\partial^{\mu}\phi
    -\frac{1}{2}\partial_{\mu}\varphi\partial^{\mu}\varphi-V(\phi,\varphi),
\end{equation}
and we include, also, ordinary matter (a comoving perfect fluid)
in the gravitational action. As in \cite{Lazkoz:2006pa} we
consider here the efective two-field potential \be V=V_0
e^{-\sqrt{6}(m\phi+n\varphi)},\label{ExponentialPot}\ee  where the
scalar field $\phi$ represents quintessence and $\varphi$
represents a phantom field. For simplicity, we assume $m>0$ and
$n>0.$

The geometry is given by the Friedmann-Robertson-Walker (FRW)
metric (in spherical coordinates): \be d s^2=-d
t^2+a(t)\left(\mathrm{d} r^2+ r^2 \mathrm{d} \theta^2+r^2
\sin^2\theta \mathrm{d} \vartheta^2 \right).\label{FRW}\ee

The field equations derived from (\ref{FRW}), are
\begin{eqnarray}
& H^2 - \sfrac16\left(\dot\phi^2-\dot\varphi^2\right)-\sfrac13 V-\sfrac13\rho_{\rm M}=0,\label{Friedmann1}\\
&\dot H=-H^2-\sfrac13\left(\dot\phi^2-\dot\varphi^2\right)+\sfrac13 V-\sfrac16 \rho_{\rm M},\label{Raych}\\
&\dot\rho_{\rm M}=-3 H\rho_{\rm M},\label{consm}\\
&\ddot\phi+3H\dot\phi-\sqrt{6}m V=0,\label{consphi}\\
&\ddot\varphi+3H\dot\varphi+\sqrt{6}n V=0,\label{consvarphi}
\end{eqnarray} where $H =\frac{\dot a(t)}{a(t)}$ denotes de Hubble expansion scalar.

The dot denotes derivative with respect the time $t.$

\section{Some results for the flat FRW case}\label{sectionIII}

In \cite{Lazkoz:2006pa} it was studied an homogenous and isotropic
universe, having dust (DM) and quintom DE using the standards
tools of the theory of dynamical systems. There are introduced the
normalized variables: $(x_\phi,\,x_\varphi,\,y)$, defined by
\begin{eqnarray} x_\phi=\frac{\dot\phi}{\sqrt{6} H},\;
x_\varphi=\frac{\dot\varphi}{\sqrt{6} H},  y=\frac{\sqrt
V}{\sqrt{3}H}.\label{vars}
\end{eqnarray}
They are related through the Friedman equation (\ref{Friedmann1}) by
$x_\phi^2-x_\varphi^2+y^2=1-\frac{\rho_{\rm M}}{3 H^2}\leq 1.$

The dynamics in a phase space is
governed by the vector field (or differential equation) \cite{Lazkoz:2006pa}:

\ben
&&x_\phi'=\frac{1}{3} \left(3 m y^2+(q-2) x_\phi\right)\label{eqxphi}\\
&&x_\vphi'=-\frac{1}{3} \left(3 n y^2-(q-2) x_\varphi \right)\label{eqxvphi}\\
&&y'=\frac{1}{3} (1+q-3(m x_\phi+n x_\varphi)) y\label{eqy} \een
defined in the phase space given by \be\Psi=\{{\bf
x}=(x_\phi,x_\vphi,y):0\le
x_\phi^2-x_\vphi^2+y^2\le1\}.\label{phase_space}\ee

Here the prime denotes differentiation with respect to a new time
variable $\tau=\log a^3,$ where $a$ is the scale factor of the
space-time. The deceleration factor $q\equiv-\ddot a a/\dot a^2$
can  be written as \be q=\frac{1}{2} \left(3
\left(x_\phi^2-x_\varphi^2-y^2\right)+1\right).\ee

\subsection{Linear analysis of non-isolated critical points}

\begin{table*}[t]
\caption[crit]{Location, existence and deceleration factor of the
critical points for $m>0$,  $n>0$ and $y>0.$ We use the notation
$\delta=m^2-n^2$ (from reference \cite{Lazkoz:2006pa}).}
\label{table1}
\begin{center}
\begin{tabular}{@{\hspace{4pt}}c@{\hspace{14pt}}c@{\hspace{14pt}}c@{\hspace{14pt}}c@{\hspace{18pt}}c@{\hspace{18pt}}c@{\hspace{2pt}}}
\hline
\hline\\[-0.3cm]
Name &$x_\phi$&$x_\vphi$&$y$&Existence&$q$\\[0.1cm]
\hline\\[-0.2cm]
$O$& $0$& $0$& $0$& All $m$ and $n$ &$\ds\frac{1}{2}$\\[0.2cm]
$C_{\pm}$ & $\pm\sqrt{1+{x_{\varphi}^*}^2}$& $x_{\varphi}^*$& $0$&
All $m$ and $n$  &$2$ \\[0.2cm]
${P}$&
$m$& $-n$ &$\sqrt{1-\delta}$&$\delta< 1$&$-1+3\delta$\\[0.2cm]
${T}$& $\ds\frac{\ds m}{\ds{2}{\delta}}$& $-\ds\frac{\ds
n}{\ds{2}{\delta}}$&
$\ds\frac{1}{2\sqrt{\delta}}$&$\delta\ge1/2$&$\ds\frac{1}{2}$\\[0.4cm]
\hline \hline
\end{tabular}
\end{center}
\end{table*}

In this section we want to investigate, in more detail, the
behavior of the two curves (hyperbolae) of non-isolated fixed
points $C_{\pm}$ with coordinates
$x_\phi=\pm\sqrt{1+{x_{\varphi}^*}^2},\,
x_\varphi=x_{\varphi}^*,\, y=0.$

From the physical viewpoint, the curves $C_\pm$ represent solutions in which the contribution
of matter and the potential energy to the total  energy density is
negligible. These solutions are therefore of stiff-fluid type
($w=1$), which in turn correspond to a decelerating universe. On
the other hand, the curves $C_\pm$ are local sources, and
therefore it is unlikely that they can represent the final stage
in the evolution of our Universe. Thus, by analyzing the sign of
the real part of the normally-hyperbolic curves $C_{\pm}$ we get
the following results (we are assuming $m>0$ and $n>0$):

\begin{enumerate}
\item If $m<n,$ $C_+$ contains an infinite arc parameterized by
${x_\varphi^*}$ such that
${x_\varphi^*}<\frac{-n-m\sqrt{1-m^2+n^2}}{m^2-n^2}$ that is a
local source. $C_-$ contains an infinite arc parameterized by
${x_\varphi^*}$ such that
${x_\varphi^*}<\frac{-n+m\sqrt{1-m^2+n^2}}{m^2-n^2}$ that is a
local source. \item If $m=n,$ $C_+$ contains an infinite arc
parameterized by ${x_\varphi^*}$ such that
${x_\varphi^*}<\frac{1-m^2}{2n}$ that is a local source. All of
$C_-$ is a local source. \item If $m>n$ there are two
possibilities
\begin{enumerate}
\item If $m^2-n^2<1,$ all of $C_-$ is a local source. A finite arc
of $C_+$ parameterized by ${x_\varphi^*}$ such that
$\frac{-n-m\sqrt{1-m^2+n^2}}{m^2-n^2}<{x_\varphi^*}<\frac{-n+m\sqrt{1-m^2+n^2}}{m^2-n^2}$
is a local source. \item If $m^2-n^2\geq 1,$ no part of $C_+$ is a
local source and all of $C_-$ is a local source.
\end{enumerate}
\end{enumerate}

In conclusion, we have that $C_-$ is always a local source
provided $m \geq n>0.$ But, we said nothing  about its global
stability. To answer this question, one needs to apply more
sophisticated tools such as the calculations of Normal Forms (we
submit the reader to section \ref{normal} for such an
investigation).

\subsection{Heteroclinic sequences}

From the local stability properties of $C_\pm, O, P$ and $T$ and
supported by several numerical integrations  \cite{Lazkoz:2006pa}
it was possible to identify heteroclinic sequences (see table
\ref{table1} for additional information concerning the critical
points $O, C_{\pm}, T, P$ reported in \cite{Lazkoz:2006pa}).

\begin{itemize}
    \item Case i a) If $0\leq m<m,$ the point $P$ is an stable node and $T$ does not exists. For this conditions
    there exist an heteroclinic sequence of type $\mathcal{K}_-\longrightarrow O\longrightarrow P$ or
    of type $\mathcal{K}_+\longrightarrow O\longrightarrow P,$ where
    $\mathcal{K}_\pm$ are infinite arcs contained in $C_\pm.$
    \item Case i b) If $0<n= m,$ the point $P$ is an stable node and $T$ does not exists. There exists an heteroclinic
    sequence of type $C_-\longrightarrow O\rightarrow P$ or of type $\mathcal{K}_+\longrightarrow O\longrightarrow P.$
    \item Case i c) If $0<n<m<\sqrt{n^2+1/2},$ the point $P$ is an stable node and $T$ does not exists. There exists an heteroclinic sequence of type $C_-\rightarrow O\rightarrow P$ or
    of type $\kappa_+\longrightarrow O\longrightarrow P,$ where
    $\kappa_+$ is a finite arc contained in $C_+.$
    \item Case ii) For $n>0$ and $\sqrt{n^2+1/2}<m\leq \sqrt{n^2+4/7},$ the point $T$ is a stable node and the point $P$ is
a saddle. For these conditions the heteroclinic sequence is of
type $C_{\pm}\longrightarrow O\longrightarrow T \longrightarrow
P,$ or of type $\kappa_+\longrightarrow O\longrightarrow T
\longrightarrow P.$
    \item Case iii) For $n>0$ and  $\sqrt{n^2+4/7}<m<\sqrt{1+n^2},$ the point T is a spiral node and the point P is a saddle. For these conditions the heteroclinic sequence is the same as in the former case.
    \item Case iv) For $n>0$ and  $m>\sqrt{1+n^2}$ the point T is a spiral node whereas the point $P$ does not exist. The
heteroclinic sequence in this case is  $C_{-}\longrightarrow
O\longrightarrow T.$
\end{itemize}

\section{Normal expansion up to arbitrary order}\label{normal}

Normal Form (NF) calculations essentially removes the quadratic,
cubic, etc., terms that are effectively neglectful up to the
desired order in the Taylor expansion around a non-hyperbolic
fixed point, e.g., $C_\pm$. The terms that cannot be removed by
using NF calculations, and they are the essential degrees of
nonlinearity (see appendix \ref{normal}). In this section we
obtain the normal form expansion for nonhyperbolic points in the
curve $C_-.$

The normal expansion up to order $N$ is given by the

\begin{prop}\label{prop4}

Let be the vector field ${\bf X}$ given by
(\ref{eqxphi}-\ref{eqy}) which is $C^\infty$ in a neighborhood of
${\bf x}^*=(x_{\phi}^*,x_{\varphi}^*,y^*)^T\in C_-.$ Let $m\geq
n>0,$ and $x_{\varphi}^*\in\mathbb{R},$ such that $\lambda_3^-=1-n
x_\phi^*+m\sqrt{1+{x_\phi^*}^2}$ is not integer, then, there exist
constants $a_r,\, b_r,\, c_r,\; r\geq 2$ (non necessarily
different from zero) and a transformation of coordinates ${\bf
x}\rightarrow {\bf y},$ such that (\ref{eqxphi}-\ref{eqy}) has
normal form

\ben
y_1'&=&\sum_{r=2}^N a_r y_1^r+{O}(|{\bf y}|^{N+1}),\label{N1}\\
y_2'&=&y_2\left(1+\sum_{r=2}^N b_r y_1^{r-1}\right)+{O}(|{\bf y}|^{N+1}),\label{N2}\\
y_3'&=&y_3\left(\lambda_3+\sum_{r=2}^N c_r
y_1^{r-1}\right)+{O}(|{\bf y}|^{N+1}),\label{N3} \een defined in
neighborhood of ${\bf y}=(0,0,0).$

\end{prop}

Proof. By applying linear coordinate transformations one is able
to reduce the linear part of ${\bf X}$ to the form $${\bf
X}_1({\bf x})= \left(\begin{array}{ccc}
0 & 0 & 0 \\
0 & 1 & 0 \\
0 & 0 & \lambda_3
\end{array}\right)\left(\begin{array}{c}
x_1\\
x_2 \\
x_3
\end{array}\right)= {\bf J}{\bf x}.$$ By the hypothesis $m\geq n>0$ we guaranteed $\lambda_3^->1$ for all
$x_{\varphi}^*\in\mathbb{R}.$ Since, the eigenvalues of ${\bf J}$
are different and ${\bf J}$ is diagonal; then, the corresponding
eigenvectors

$$B=\left\{x_1^{m_1} x_2^{m_2} x_3^{m_3}{\bf e}_i | m_j\in\mathbb{N}, \sum m_j=r, i,j=1,2,3\right\}$$
form a basis of $H^r.$ Since
$${\bf L}_{\bf J} {\bf x}^{\bf m}{\bf e}_i
=\left\{({\bf m}\cdot {\bf \lambda})-\lambda_i\right\}
{\bf x}^{\bf m}{\bf e}_i,$$ the eigenvectors in $B$ for which
$\Lambda_{{\bf m},i}\equiv ({\bf m}\cdot {\bf \lambda})-\lambda_i\neq 0$
form a basis of $B^r={\bf L}_{\bf J}(H^r).$ The eigenvectors associated to the resonant
eigenvalues, i.e., those such that $\Lambda_{{\bf m},i}=0,$ form a basis for the
complementary subspace, $G^r,$ of $B^r$ in $H^r.$

Since $\lambda_1=0,$ the resonant equations of order $r$ (with
$m_1+m_2+m_3=r$) has unique solution

\ben
m_2+\lambda_3 m_3=0 \Rightarrow m_1=r, m_2=m_3=0,\\
m_2+\lambda_3 m_3=1 \Rightarrow m_1=r-1, m_2=1, m_3=0,\\
m_2+\lambda_3 m_3=\lambda_3 \Rightarrow m_1=r-1, m_2=0, m_3=1
\label{hom3}. \een

Equation (\ref{hom3}) has another solution given by
$m_1=r-\lambda_3, m_2=\lambda_3, m_3=0,$ provided $\lambda_3$ is
integer satisfying $1<\lambda_3\leq r.$ This solution is discarded
by our hypothesis on $\lambda_3.$

Then, $\left\{x_1^r {\bf e}_1, x_1^{r-1} x_2 {\bf e}_2,x_1^{r-1}
x_3 {\bf e}_3\right\},$ form a basis for  $G^r$ in $H^r.$

By applying theorem \ref{NFTheorem}, we have that there exists a
coordinate transformation ${\bf x}\rightarrow {\bf y},$ such that
(\ref{eqxphi}-\ref{eqy}) has normal form (\ref{N1}-\ref{N3}) where
$a_r, b_r$ and $c_r$ are some real constants. $\blacksquare$

The values of of all these constants can be uniquely determined,
up to the desired order, by inductive construction in $r.$ The
idea is to introduce polynomial coordinate transformations ${\bf
x} \rightarrow {\bf x}+ {\bf h}_r({\bf x})$ such that ${\bf
L}_{\bf J}{\bf h}_r ({\bf x})={\bf X}_r({\bf x})+\text{higher order terms}.$  The higher
order terms of the expansion are modified by successive
coordinates transformations. Suppose we apply the $r$-th
transformation, for a fixed $r.$ Then, all non-resonant terms of
order $r,$ are eliminated. The $r$-th transformation modifies the
expansion terms of order higher than $r,$ but resonant terms of
order less than $r$ are not affected. One then removes the
non-resonant terms of order $r+1$ by introducing the $(r+1)$-th
transformation, and so on. In each step one obtains the desired
coefficients. Finally, we can truncated the expansion up to order
$N.$

We see that (\ref{N1}-\ref{N3}) satisfy the conditions of the existence and uniqueness theorem
of differential equations. Then, there exist a unique solution of (\ref{N1}-\ref{N3})
passing through $(y_{10},y_{20},y_{30}),$ at $t=0.$ By neglecting the error terms one is
able to integrate the resulting approximated system

\ben
y_1'&=&\sum_{r=2}^N a_r y_1^r,\label{approxN1}\\
y_2'&=&y_2\left(1+\sum_{r=2}^N b_r y_1^{r-1}\right),\label{approxN2}\\
y_3'&=&y_3\left(\lambda_3+\sum_{r=2}^N c_r y_1^{r-1}\right),\label{approxN3} \een
with initial condition $\left(y_1(t_0),y_2(t_0),y_3(t_0)\right)=(y_{10},y_{20},y_{30}).$

The general solution of (\ref{approxN1}-\ref{approxN3}) is as follows.

If $y_{10}=0,$ then $y_1(t)=0,$  $y_2(t)=y_{20}e^{\tau-\tau_0}$ and
$y_3(t)=y_{30}e^{\lambda_3\left(\tau-\tau_0\right)}$ for all
$t\in \mathbb{R}.$ Then, the orbit approach the origin as $\tau\rightarrow-\infty$ provided $\lambda_3>0.$

If $y_{10}\neq 0,$ then (\ref{approxN1}-\ref{approxN3}) can be integrated in quadratures as

\ben
\tau-\tau_0=\int_{y_{10}}^{y_1} \left(\sum_{r=2}^N a_r \zeta^r\right)^{-1} \mathrm{d}\zeta,\label{tauy1}\\
y_2(t)=y_{20} e^{\tau-\tau_0} \prod_{r=2}^N \exp\left[b_r\int_{\tau_0}^\tau y_1(t)^{r-1} \mathrm{d} t\right],\label{eqNy2}\\
y_3(t)=y_{30}e^{\lambda_3\left(\tau-\tau_0\right)} \prod_{r=2}^N
\exp\left[c_r\int_{\tau_0}^\tau y_1(t)^{r-1} \mathrm{d}
t\right].\label{eqNy3} \een

The $y_1$-component of the orbit passing through
$(y_{10},y_{20},y_{30})$ at $\tau=\tau_0$ with $y_{10}\neq 0$ is
obtained by inverting the quadrature (\ref{tauy1}).

The other components are given by

\ben
y_2=y_{20} \exp\left[\int_{y_{10}}^{y_1}\frac{1+\sum_{r=2}^N b_r \zeta^{r-1}}{\sum_{r=2}^N a_r \zeta^r}\mathrm{d} \zeta\right],\\
y_3=y_{30} \exp\left[\int_{y_{10}}^{y_1}\frac{1+\sum_{r=2}^N c_r
\zeta^{r-1}}{\sum_{r=2}^N a_r \zeta^r} \mathrm{d} \zeta\right].
\een This orbit does not admit, in general, a prolongation to all
the time values. If the maximal interval of definition, $(\alpha,
\beta)$ of the solution $y_1$ is such that $\alpha$ is finite,
then the orbits diverges in the $y_1$-direction as
$\tau\rightarrow \alpha^+.$

\subsection{Treatable case: normal expansion to third order for
$C_-$}

In this section we show normal form expansions for the vector
field (\ref{eqxphi}-\ref{eqy}) defined in a vicinity of $C_-$
expressed in the form of proposition \ref{Prop2.5}. The proof is
given in three steps. First, by applying linear coordinate
transformations (translation to the origin and similarity
transformation to reduce the Jacobian matrix to its real Jordan
form) we reduce the linear part in its simpler form. Second, we
perform a quadratic coordinate transformation in order to reduce
non-resonant second order terms. Third, we perform a cubic
coordinate transformation that allows to eliminate non-resonant
terms of third order. In principle it is possible to eliminate all
the non-resonant terms of all orders. However, we find the system
computationally treatable to third order.

\begin{prop}\label{Prop2.5} Let be the vector field ${\bf X}$ given by
(\ref{eqxphi}-\ref{eqy}) which is $C^\infty$ in a neighborhood of
${\bf x}^*=(x_{\phi}^*,x_{\varphi}^*,y^*)^T\in C_-.$ Let $m\geq
n>0,$ and $x_{\varphi}^*\in\mathbb{R},$ such that $\lambda_3^-=1-n
x_\phi^*+m\sqrt{1+{x_\phi^*}^2}$ is not integer, then, there exist
a transformation to new coordinates $x\rightarrow z,$ such that
(\ref{eqxphi}-\ref{eqy}), defined in a vicinity of ${\bf x}^*,$
has normal form

\ben \label{2_17} z_1'&=&O(|z |^4),
\\
\label{2_18} z_2' &=& z_2+O(|z |^4),
\\
\label{2_19} z_3'&=&\left(\lambda^-_3+c_2 z_1+c_3
z_1^2\right)z_3+O(|z |^4),\een where  $c_2=-n+\frac{m
x_\varphi^*}{\sqrt{1+{x_\varphi^*}^2}}$ and $c_3=-\frac{n
x_\varphi^*}{2\left(1+{x_\varphi^*}^2\right)}+\frac{m}{2\sqrt{1+{x_\varphi^*}^2}}.$

\end{prop}

Proof.

By using linear coordinates transformations the system
(\ref{eqxphi}-\ref{eqy}) is reduced to \be {\bf x}'={\bf J} {\bf
x}+{\bf X}_2({\bf x})+{\bf X}_3({\bf x})\label{Jordan3}\ee
where ${\bf x}$ stands for the phase vector ${\bf x}=\left(x_1,\,x_2,\,x_3\right)^T,$ ${\bf J}$ stands for the Jordan Form of the matrix of derivatives \be {\bf J}=\left(\begin{array}{ccc} 0 & 0 & 0\\
                0 & 1 & 0\\
                0 & 0 & 1-n {x_{\varphi}^*}+m \sqrt{1+{x_{\varphi}^*}^2}\end{array}\right)\label{JordanMatrix}\ee
\be {\bf X}_2({\bf x})=\left(\begin{array}{c} X_{(1,1,0),1}{x_1
   \,x_2}+X_{(0,0,2),1} x_3^2\vspace{10pt}\\
               X_{(2,0,0),2} x_1^2+ X_{(0,2,0),2} x_2^2+ X_{(0,0,2),2} x_3^2 \vspace{10pt}\\
                X_{(1,0,1),3} x_1  x_3+   X_{(0,1,1),3} x_2  x_3\end{array}\right),\label{X2}\ee
                where the coefficients ${\bf X}_{{\bf m},i}$ are displayed in table \ref{tab1} for
                the allowed ${\bf m}.$

\begin{table*}[t!]\caption[coef]{Coefficients of the vector field ${\bf X}_2(\bf{x})$.}
\begin{tabular}{@{\hspace{4pt}}c@{\hspace{14pt}}c@{\hspace{14pt}}c@{\hspace{14pt}}c@{\hspace{2pt}}}
\hline
\hline\\[-0.3cm]
${\bf m}$ &$X_{{\bf m},1}$&$X_{{\bf m},2}$&$X_{{\bf m},3}$\\[0.1cm]
\hline\\[-0.2cm]
$(2,0,0)$ & $0$ & $-\frac{{x_{\varphi}^*}}{2 \left({x_{\varphi}^*}^2+1\right)}$ & $0$ \\[0.3cm]
$(1,1,0)$ & $\frac{1}{{x_{\varphi}^*}}$ & $0$ & $0$ \\[0.3cm]
$(1,0,1)$ & $0$ & $0$ & $-n+\frac{m {x_{\varphi}^*}}{\sqrt{1+{x_{\varphi}^*}^2}}$ \\[0.3cm]
$(0,2,0)$ & $0$ & $\frac{3}{2 {x_{\varphi}^*}}$ & 0\\[0.3cm]
$(0,1,1)$ & $0$ & $0$ & $-\frac{
   \left(-m \sqrt{1+{x_{\varphi}^*}^2}+n {x_{\varphi}^*}-1\right)}{{x_{\varphi}^*}}$ \\[0.3cm]
$(0,0,2)$ & $m  {x_{\varphi}^*}\sqrt{1+{x_{\varphi}^*}^2}-n
\left(1+{x_{\varphi}^*}^2\right)$ & $\frac{1}{2} {x_{\varphi}^*}
\left(-2 m
   \sqrt{1+{x_{\varphi}^*}^2}+2 n {x_{\varphi}^*}-1\right)$ & $0$ \\[0.4cm]
\hline \hline
\end{tabular}\label{tab1}
\end{table*}

The third order terms become:

\be{\bf X}_3({\bf x})=\left(\begin{array}{c} X_{(3,0,0),1}{x_1^3}+X_{(1,2,0),1} x_1 x_2^2+X_{(1,0,2),1} x_1 x_3^2\vspace{10pt}\\
               X_{(2,1,0),2} x_1^2 x_2+ X_{(0,3,0),2} x_2^3 +X_{(0,1,2),2} x_2 x_3^2\vspace{10pt}\\
                X_{(2,0,1),3} x_1^2 x_3 +  X_{(0,2,1),3} x_2^2 x_3 +  X_{(0,0,3),3} x_3^3
                \end{array}\right),\label{X3}\ee with the
                coefficients defined as in table \ref{tab3}.

\begin{table*}[t!]\caption[coef]{Coefficients of the vector field ${\bf X}_3(\bf{x})$.}
\begin{tabular}{@{\hspace{4pt}}c@{\hspace{14pt}}c@{\hspace{14pt}}c@{\hspace{14pt}}c@{\hspace{2pt}}}
\hline
\hline\\[-0.3cm]
${\bf m}$ &$X_{{\bf m},1}$&$X_{{\bf m},2}$&$X_{{\bf m},3}$\\[0.1cm]
\hline\\[-0.2cm]
$(3,0,0)$ & $-\frac{1}{2 \left({x_\varphi^*}^2+1\right)}$ & $0$ & $0$ \\[0.3cm]
$(2,1,0)$ & $0$ & $-\frac{1}{2 \left({x_\varphi^*}^2+1\right)}$ & $0$ \\[0.3cm]
$(2,0,1)$ & $0$ & $0$ & $-\frac{1}{2 \left({x_\varphi^*}^2+1\right)}$ \\[0.3cm]
$(1,2,0)$ & $\frac{1}{2 {x_\varphi^*}^2}$ & $0$ & $0$ \\[0.3cm]
$(1,0,2)$ & $-\frac{1}{2}$ & $0$ & $0$ \\[0.3cm]
$(0,3,0)$ & $0$ & $\frac{1}{2 {x_\varphi^*}^2}$ & $0$ \\[0.3cm]
$(0,2,1)$ & $0$ & $0$ & $\frac{1}{2 {x_\varphi^*}^2}$ \\[0.3cm]
$(0,1,2)$ & $0$ & $-\frac{1}{2}$ & $0$ \\[0.3cm]
$(0,0,3)$ & $0$ & $0$ & $-\frac{1}{2}$ \\[0.4cm]
\hline \hline
\end{tabular}\label{tab3}
\end{table*}

\textbf{Second step: simplifying the quadratic part}

\begin{table*}[t!]\caption[coef]{Eigenvalues of ${\bf L}^(2)_{\bf J}: H^2\rightarrow H^2$.}
\begin{tabular}{@{\hspace{4pt}}c@{\hspace{14pt}}c@{\hspace{14pt}}c@{\hspace{14pt}}c@{\hspace{2pt}}}
\hline
\hline\\[-0.3cm]
${\bf m}$ &$\Lambda_{{\bf m},1}$&$\Lambda_{{\bf m},2}$&$\Lambda_{{\bf m},3}$\\[0.1cm]
\hline\\[-0.2cm]
$(2,0,0)$ & - & $-1$ & - \\[0.3cm]
$(1,1,0)$ & $1$ & - & - \\[0.3cm]
$(1,0,1)$ & - & - & $0$ \\[0.3cm]
$(0,2,0)$ & - & $1$ & - \\[0.3cm]
$(0,1,1)$ & - & - & $1$ \\[0.3cm]
$(0,0,2)$ & $2\left(1-n {x_{\varphi}^*}+m \sqrt{1+{x_{\varphi}^*}^2}\right)$ & $1-2 n {x_{\varphi}^*}+2 m \sqrt{1+{x_{\varphi}^*}^2}$ & - \\[0.4cm]
\hline \hline
\end{tabular}\label{tab2}
\end{table*}

By the hypotheses the eigenvalues
$\lambda_1=0,\lambda_2=1,\lambda_3^-=1-n {x_{\varphi}^*}^2+m
\sqrt{1+{x_{\varphi}^*}^2}$ of ${\bf J}$ are different. Hence, its
eigenvectors form a basis of $\mathbb{R}^3.$ The linear operator
$${\bf L}^{(2)}_{\bf J}: H^2\rightarrow H^2$$ has eigenvectors ${\bf
x}^{\bf m}{\bf e}_i$ with eigenvalues $\Lambda_{{\bf
m},i}=m_1\lambda_1+m_2\lambda_2+\lambda_3 m_3-\lambda_i,$
$i=1,2,3,$ $m_1,m_2,m_3\geq 0,$ $m_1+m_2+m_3=2.$ The eigenvalues
$\Lambda_{{\bf m},i}$ for the available ${\bf m},i$ are displayed
in table \ref{tab2}. To obtain the normal form of
(\ref{2_17}-\ref{2_19}) wee must look for resonant terms, i.e.,
those terms of the form ${\bf x}^{\bf m}{\bf e}_i$ with ${\bf m}$
and $i$ such that $\Lambda_{{\bf m},i}=0$ for the available ${\bf
m},i.$ Only one term is resonant of second order:
$\Lambda_{(1,0,1),3}=0\rightarrow c_2 y_1 y_3 {\bf e}_3.$

The required function $${\bf h}_2: H^2\rightarrow H^2$$ to
eliminate the non-resonant quadratic terms is given by \be {\bf
h}_2({\bf y})=\left(\begin{array}{c}
\frac{X_{(1,1,0),1}}{\Lambda_{(1,1,0),1}}{y_1
   \,y_2}+\frac{X_{(0,0,2),1}}{\Lambda_{(0,0,2),1}} y_3^2\vspace{10pt}\\
               \frac{X_{(2,0,0),2}}{\Lambda_{(2,0,0),2}} y_1^2+ \frac{X_{(0,2,0),2}}{\Lambda_{(0,2,0),2}} y_2^2+ \frac{X_{(0,0,2),2}}{\Lambda_{(0,0,2),2}} y_3^2 \vspace{10pt}\\
                 \frac{X_{(0,1,1),3}}{\Lambda_{(0,1,1),3}} y_2  y_3\end{array}\right),\label{h2}\ee

The quadratic transformation \be {\bf x}\rightarrow {\bf y}+{\bf
h}_2(\bf y)\label{qtransform}\ee with ${\bf h}_2$ defined as in
(\ref{h2}) is the coordinate transformation required in theorem
\ref{NFTheorem}. By applying this theorem we prove the existence
of the required constant $c_2.$

To finish the proof, let us calculate the value of $c_2.$

By applying the transformation (\ref{qtransform}) the vector field
(\ref{Jordan3}) transforms to \be {\bf y}'={\bf J y}-{\bf
L}^{(2)}_{\bf J} {\bf h}_2 ({\bf y})+{\bf X}_2(\bf y)+\tilde{{\bf
X}}_3(\bf y)+{O}(|{\bf y}|^4),\label{Jordan4}\ee

Since \be -{\bf L}^{(2)}_{\bf J} {\bf h}_2 ({\bf y})+{\bf
X}_2({\bf y})= X_{(1,0,1),3}y_1 y_3 {\bf e}_3,\ee we have \be {\bf
y}'={\bf J y}+X_{(1,0,1),3}y_1 y_3 {\bf e}_3+\tilde{{\bf X}}_3(\bf
y)+{O}(|{\bf y}|^4)\ee , i.e., $c_2=X_{(1,0,1),3}=-n+\frac{m
{x_{\varphi}^*}}{\sqrt{1+{{x_{\varphi}^*}}^2}}.$

The vector field $\tilde{{\bf X}}_3(\bf y)$ introduced above has
the coefficients:

\begin{widetext}

\ben
{\tilde{X}}_{{ (2,0,1),3}}=\frac{m}{2
\sqrt{{x_\varphi^*}^2+1}}-\frac{n {x_\varphi^*}}{2  \left({x_\varphi^*}^2+1\right)},\nonumber\\
   {\tilde{X}}_{{ (1,2,0),1}}=\frac{3}{{x_\varphi^*}^2}
   ,\nonumber\\
   {\tilde{X}}_{{ (1,0,2),1}}=-\frac{n^2 \delta ^2+m\left(\delta +m\left(\delta ^2-1\right)\right)-n
   {x_\varphi^*}[2 \text{m$\delta $}+1]+1}{\left(\lambda
   _3\right){}^-}
   ,\nonumber\\
   {\tilde{X}}_{{ (1,0,2),2}}=\frac{{x_\varphi^*}\left[2 {x_\varphi^*}
   m^2+\frac{\left({x_\varphi^*}-2 \left(2 \left(\delta
   ^2-1\right) n+n\right)\right) m}{\delta }+n\left(2 n {x_\varphi^*}-1\right)\right]}
   {2 {\lambda_3^-}},\nonumber\\
      {\tilde{X}}_{{ (0,3,0),2}}=\frac{5}{{x_\varphi^*}^2},\nonumber\\
      {\tilde{X}}_{{ (0,2,1),3}}=\frac{\left(-\text{m$\delta $}+n {x_\varphi^*}-2\right)
   \left(-2 \text{m$\delta $}+2 n {x_\varphi^*}-3\right)}
   {2 {x_\varphi^*}^2},\nonumber\\
   {\tilde{X}}_{{
(0,1,2),1}}=\frac{\delta \left(4 \delta ^2 {x_\varphi^*} m^3+4
   \delta  \Delta _1 m^2+\Delta _2 m+\text{n$\delta $} \Delta
   _3\right)}{2 {\lambda_3^-} {x_\varphi^*}},\nonumber \\
   {\tilde{X}}_{{ (0,1,2),2}}=-2 \left(\delta ^2-1\right) n^2+{x_\varphi^*}[4
   \text{m$\delta $}+3] n-\text{m$\delta $}(2 \text{m$\delta $}+3)-3,\nonumber\\
   {\tilde{X}}_{{
(0,0,3),3}}=-4 n {x_\varphi^*}\left[n{x_\varphi^*}-2\right]-5,
\een

\end{widetext}

where $\delta =\sqrt{{x_\varphi^*}^2+1},$ ${\Delta }_{{ 1}}=2
{x_\varphi^*}-n\left(3 {x_\varphi^*}^2+1\right),$ ${\Delta }_{{
2}}=4 n\left(-4 {x_\varphi^*}^2+n\left(3
   {x_\varphi^*}^2+2\right)
   {x_\varphi^*}-2\right)+5 {x_\varphi^*},$ ${\Delta }_{{ 3}}=-4 n {x_\varphi^*}
   \left[n {x_\varphi^*}-2\right]-5.$

\textbf{Third step: simplifying the cubic part}

\begin{table*}[t!]\caption[coef]{Eigenvalues of ${\bf L}^{(3)}_{\bf J}: H^3\rightarrow H^3$.}
\begin{tabular}{@{\hspace{4pt}}c@{\hspace{14pt}}c@{\hspace{14pt}}c@{\hspace{14pt}}c@{\hspace{2pt}}}
\hline
\hline\\[-0.3cm]
${\bf m}$ &$\Lambda_{{\bf m},1}$&$\Lambda_{{\bf m},2}$&$\Lambda_{{\bf m},3}$\\[0.1cm]
\hline\\[-0.2cm]
$(2,0,1)$ & - & - & $0$ \\[0.3cm]
$(1,2,0)$ & $2$ & - & - \\[0.3cm]
$(1,0,2)$ & $2 \left(m\sqrt{1+{x_\varphi^*}^2}-n
   {x_\varphi^*}+1\right)$ & $2 m\sqrt{1+{x_\varphi^*}^2}-2 n
   {x_\varphi^*}+1$ & - \\[0.3cm]
$(0,3,0)$ & - & $2$ & - \\[0.3cm]
$(0,2,1)$ & - & - & $2$ \\[0.3cm]
$(0,1,2)$ & $2 m\sqrt{1+{x_\varphi^*}^2}-2 n
   {x_\varphi^*}+3$ & $2 \left(m\sqrt{1+{x_\varphi^*}^2}-n
   {x_\varphi^*}+1\right)$ & - \\[0.3cm]

$(0,0,3)$ & - & - & $2 \left(m\sqrt{1+{x_\varphi^*}^2}-n
   {x_\varphi^*}+1\right)$
\\[0.4cm]\hline
\hline
\end{tabular}\label{tab4}
\end{table*}

After the last two steps, the equation (\ref{Jordan3}) is
transformed to \be {\bf y}'={\bf J y}+c_2 y_1 y_3 {\bf
e}_3+\tilde{{\bf X}}_3({\bf y})+{\bf O}(|{\bf y}|^4).\ee

There is only one term of order three which is resonant (see table
\ref{tab4}): $\Lambda_{(2,0,1),3}=0\rightarrow c_3 z_1^2 z_3 {\bf
e}_3.$

As in the last step, in order to eliminate non-resonant terms of
third order we will consider the coordinate transformation ${\bf
y}\rightarrow {\bf z}$ given by \be {\bf y}={\bf z}+{\bf h}_3
({\bf z})\label{transfz}\ee where
$${\bf h}_3: H^3\rightarrow
H^3$$  is defined by \be {\bf h}_3({\bf z})=\left(
\begin{array}{l}
\frac{{\tilde{X}}_{{\rm (1,2,0),1}}}{{\Lambda }_{{\rm (1,2,0),1}}}z_{{\rm 1}}z^{{\rm 2}}_{{\rm 2}}{\rm +}\frac{{\tilde{X}}_{{\rm (1,0,2),1}}}{{\Lambda }_{{\rm (1,0,2),1}}}z_{{\rm 1}}z^{{\rm 2}}_{{\rm 3}}{\rm +}\frac{{\tilde{X}}_{{\rm (0,1,2),1}}}{{\Lambda }_{{\rm (0,1,2),1}}}z_{{\rm 2}}z^{{\rm 2}}_{{\rm 3}} \\
\frac{{\tilde{X}}_{{\rm (1,0,2),2}}}{{\Lambda }_{{\rm (1,0,2),2}}}z_{{\rm 1}}z^{{\rm 2}}_{{\rm 3}}{\rm +}\frac{{\tilde{X}}_{{\rm (0,3,0),2}}}{{\Lambda }_{{\rm (0,3,0),2}}}z^{{\rm 3}}_{{\rm 2}}{\rm +}\frac{{\tilde{X}}_{{\rm (0,1,2),2}}}{{\Lambda }_{{\rm (0,1,2),2}}}z_{{\rm 2}}z^{{\rm 2}}_{{\rm 3}} \\
\frac{{\tilde{X}}_{{\rm (0,2,1),3}}}{{\Lambda }_{{\rm
(0,2,1),3}}}z^{{\rm 2}}_{{\rm 2}}z_{{\rm 3}}{\rm
+}\frac{{\tilde{X}}_{{\rm (0,0,3),3}}}{{\Lambda }_{{\rm
(0,0,3),3}}}z^{{\rm 3}}_{{\rm 3}} \end{array}
\right),\label{h3}\ee where $\Lambda_{{\bf
                m},i}$ are the eigenvalues of the operator linear operator
$${\bf L}^{(3)}_{\bf J}: H^3\rightarrow H^3$$ associated to the
eigenvectors ${\bf x}^{{\bf m}}{\bf e}_i.$ In table \ref{tab4} are
shown these eigenvalues. The associated eigenvectors form a basis
of $H^3$ (the space of vector fields with polynomial components of
third degree) because the eigenvalues of ${\bf J}$ are different.

The transformation (\ref{transfz}) is the required by theorem
(\ref{NFTheorem}). By using this theorem we prove the existence of
the required constant $c_3.$ To find its we must to calculate
$$-{\bf L}^{(3)}_{\bf J} {\bf h}_3({\bf z})+\tilde{{\bf X}}_3({\bf z}),$$ which it is equal to
$$\tilde{X}_{(2,0,1),3} z_1^3 {\bf e}_3$$
where ${\bf e}_i,\; i=1,2,3,$ is the canonical basis in
$\mathbb{R}^n.$ Then,
$$c_3=\frac{m}{2 \sqrt{1+{x_\varphi^*}^2}}-\frac{n {x_\varphi^*}}{2
   \left({x_\varphi^*}^2+1\right)}.$$

Observe that the transformation ${\bf h}_3$ does not affect the
value of the coefficient of the resonant term of order $r=2.$
Then, the result of the proposition follows. $\blacksquare$

\subsubsection{Unstable manifold to third order}

For $\lambda_3^->0$, the origin has a 2-dimensional local unstable
manifold tangent to the plane $z_2$-$z_3$ at $\bf 0$ given by \ben
W^u_{{loc}}({\mathbf 0})=\{\left(z_1,z_2,z_3\right) \in
\mathbb{R}^3: z_1=h(z_2,z_2), \nonumber\\ {\bf Dh}({\mathbf
0})={\mathbf 0}, \left|\left(z_2,z_3\right)^T\right|<\delta \},
\een where $h:{\mathbb R}^{2}\rightarrow {\mathbb R}$ is a $C^r$
function and $\delta >0$ is small enough.

Using the invariance of $W^u_{{loc}}({\mathbf 0})$ under the
dynamics of (\ref{2_17}-\ref{2_19}) we obtain a quasi-linear
partial differential equation that $h$ must satisfy:

\be\mathcal{N}\left(h\left(z_2,z_3\right)\right)={\bf O}(|{\bf
z}|^4)\ee where we have defined the differential operator

\be \mathcal{N}\left(h\left(z_2,z_3\right)\right)\equiv z_2
\frac{\partial h}{\partial z_2}+ z_3 \left(c_3 h^2+c_2
   h+\lambda _3^-\right)
   \frac{\partial h}{\partial z_3}.\ee This system is solved up to
   order ${\bf O}(|{\bf z}|^4).$

Assuming that $W_{\text{loc}}^c({\bf 0})$ is
$C^4\left(\mathbb{R}^3\right)$ we can expressed it as the graph

\ben h\left(z_2,z_3\right)\equiv h_{3,0} z_2^3+h_{2,0} z_2^2+z_3
h_{2,1} z_2^2+z_3
   h_{1,1} z_2+\nonumber\\z_3^2 h_{1,2} z_2+z_3^2 h_{0,2}+z_3^3 h_{0,3}+{\bf O}(|{\bf
   z}|^4).\een This function satisfy the tangentiality conditions $h(0,0)=\partial_{z_2} h (0,0)=\partial_{z_3}  h (0,0)=(0,0).$

By substitution in the differential equation
$\mathcal{N}\left(h\left(z_2,z_3\right)\right)=0$ and by
discarding the error terms we find

\begin{widetext}

\be 3 h_{3,0} z_2^3+2 h_{2,0} z_2^2+h_{2,1}
z_3\left({\lambda_3^-}+2\right) z_2^2+h_{1,1}
z_3\left({\lambda_3^-}+1\right) z_2+h_{1,2} z_3^2\left[2
{\lambda_3^-}+1\right] z_2+2 z_3^2 h_{0,2} {\lambda_3^-}+3 z_3^3
h_{0,3} {\lambda_3^-}=0 \label{expansion}\ee
\end{widetext}
Using the condition $\lambda_3^->0$ we get
$h_{3,0}=h_{2,1}=h_{2,0}=h_{1,2}=h_{1,1}=h_{0,3}=h_{0,2}=0.$ Then,
the unstable manifold of the origin is, up to order ${O}(|{\bf
   z}|^4)$,
$W^u_{{loc}}({\mathbf 0})=\{\left(z_1,z_2,z_3\right) \in
\mathbb{R}^3: z_1=0, z_2^2+z_3^2<\delta^2\}$ where $\delta$ is a
real value small enough. Therefore, the dynamics of
(\ref{2_17}-\ref{2_19}) restricted to the unstable manifold, is
given, up to order ${\bf O}(|{\bf
   z}|^4),$ by
$z_1\equiv 0,\, z_2(\tau )=e^{\tau } z_{20},\, z_2(\tau
)=e^{\lambda_3^-\tau } z_{30},$ where
$z_{20}^2+z_{30}^2<\delta^2.$ This means that $\lim_{\tau \to
-\infty } \left(z_1(\tau ),
   \, z_2(\tau ),\, z_3(\tau
   )\right)=\left(0,0,0\right).$ Then, the origin is the past attractor
for an open set of orbits of (\ref{2_17}-\ref{2_19}). In the
original coordinates this means that the past asymptotic dynamics
of an open set of orbits of the system can be accurately
approximated by a massless quintom model located at $C_-.$ Let us
investigate the center manifold.

\subsubsection{Center manifold to third order}

For $\lambda_3^->0$, the origin has a 1-dimensional local unstable
manifold tangent to the plane $z_1$ at $\bf 0$ given by \ben
W^c_{{loc}}({\mathbf 0})=\{\left(z_1,z_2,z_3\right) \in
\mathbb{R}^3: z_2=f(z_1), z_3=g(z_1), \nonumber\\ {Df}(0)=0,
{Dg}(0)=0, \left|z_1\right|<\delta \}, \een where $f:{\mathbb
R}\rightarrow {\mathbb R}$ and $g:{\mathbb R}\rightarrow {\mathbb
R}$ are $C^r$ functions of $z_1$ and $\delta
>0$ is small enough.

Using the invariance  de $W^c_{{loc}}({\mathbf 0})$ under the
dynamics of (\ref{2_17}-\ref{2_19}) we obtain a system of two
ordinary differential equations that $f$ and $g$ must satisfy:
\ben
f(z_1)-f'(z_1) {O}\left({|{\bf z}|^4}\right)={O}\left({|{\bf z}|^4}\right), \label{2_42a}\\
\left(\lambda_3^-+c_2 z_1+c_3
z_1^2\right)g(z_1)-g'(z_1){O}\left({|{\bf
z}|^4}\right)={O}\left({|{\bf z}|^4}\right). \label{2_42b}
 \een

In order to obtain an expression for the center manifold of the
origin up to order ${\bf O}\left({|{\bf z}|^4}\right)$ we need to
solve equations (\ref{2_42a}-\ref{2_42b}) up to the same order.
Assuming that $W^c_{{loc}}({\mathbf 0})$ is $C^4,$ it can be
expressed as

\ben
f(z_1)=a z_1^2+b z_1^3+O\left({|z_1|^4}\right), \label{2_43a}\\
g(z_1)=c z_1^2+d z_1^3+O\left({|z_1|^4}\right). \label{2_43b}
 \een
The representation (\ref{2_43a}-\ref{2_43b}) satisfy the
tangentiality conditions $f(0)=g(0)=f'(0)=g'(0).$ By substitution
of (\ref{2_43a}-\ref{2_43b}) in (\ref{2_42a}-\ref{2_42b}) and
neglecting error terms we get

\ben
&&a z_1^2+b z_1^3=O\left({|z_1|^4}\right)\implies a=b=0, \label{2_44a}\\
&&\left(c c_2+d {\lambda_3^-}\right) z_1^3+c {\lambda_3^-}
z_1^2=O\left({|z_1|^4}\right) \nonumber\\
&&\implies c {\lambda_3^-}=c c_2+d {\lambda_3^-}=0
\nonumber\\
&&\implies c =d =0. \label{2_44b}
 \een The last implication follows from the fact that $\lambda_3^->0.$
Then, the center manifold of the origin is, up to order ${O}(|{\bf
   z}|^4)$,
$W^c_{{loc}}({\mathbf 0})=\{\left(z_1,z_2,z_3\right) \in
\mathbb{R}^3: z_2=z_3=0, |z_1|<\delta\}$ where $\delta$ is a real
value small enough.

Let us describe the center manifold, up to the prescribed order,
as a graph in the original variables.

We have

\begin{align} & x_1\equiv x_{\phi
}+\sqrt{{x_\varphi^*}^2+1}=-\frac{z_1\left(z_1+2
{x_\varphi^*}\right)}{2
\sqrt{{x_\varphi^*}^2+1}}+O\left({|z_1|^4}\right)\label{2_45},\\
& x_2\equiv x_{\varphi}-{x_\varphi^*}=\frac{{x_\varphi^*}
z_1^2}{2 {x_\varphi^*}^2+2}+z_1+O\left({|z_1|^4}\right)\label{2_46},\\
& x_3\equiv y=O\left({|z_1|^4}\right)\label{2_47}. \end{align}

By taking the inverse, up to fourth order, of (\ref{2_46}) we have
the expression for $z_1$:

\be z_1=\frac{{x_\varphi^*}^2 x_2^3}{2
   \left({x_\varphi^*}^2+1\right)^2}-\frac{{x_\varphi^*} x_2^2}
   {2 \left({x_\varphi^*}^2+1\right)}+x_2+O\left({|x_2|^4}\right)\label{2_48}.\ee
Substituting (\ref{2_48}) in (\ref{2_45}-\ref{2_47}) we have that
the center manifold of the origin is given by the graph:
$\{(x_1,x_2,x_3)\in\mathbb{R}^3: x_1=\frac{{x_\varphi^*} x_2^3}{2
   \left({x_\varphi^*}^2+1\right){}^{5/2}}-\frac{x_2^2}{2
   \left({x_\varphi^*}^2+1\right){}^{3/2}}-
   \frac{{x_\varphi^*}
   x_2}{\sqrt{{x_\varphi^*}^2+1}}+O\left({|x_2|^4}\right), x_3=O\left({|x_2|^4}\right),
   |x_2|<\delta\}$ for $\delta>0$ small enough.

In conclusion, the analysis up to fourth order does not allows to
classify the global past asymptotic dynamics of quintom model.
Since the numerical experiments in \cite{Lazkoz:2006pa} suggest
that there is an open set of orbits that tends to infinity, then
it is worthy to investigate the dynamics at infinity. 

\section{Analysis at infinity}\label{infinite}

\begin{table*}[t]
\caption[crit]{Location and existence conditions for the critical
points at infinity.} \label{critatinfinity}
\begin{center}
\begin{tabular}{@{\hspace{4pt}}c@{\hspace{14pt}}c@{\hspace{18pt}}c@{\hspace{18pt}}c@{\hspace{2pt}}}
\hline
\hline\\[-0.3cm]
Name &$\theta_1$&$\theta_2$&Existence \\[0.1cm]
\hline\\[-0.2cm]
$P_1^\pm$& $0$& $\pm\frac{\pi}{2}$& always\\[0.2cm]
$P_2^\pm$ & $\pi$ & $\pm\frac{\pi}{2}$&  always\\[0.2cm]
$P_3^\pm$& $\frac{\pi}{4}$& $\pm \cos
^{-1}\left(-\frac{m}{n}\right)$ &$-\pi<\pm\cos
^{-1}\left(-\frac{m}{n}\right)\leq \pi, n\neq 0$\\[0.2cm]
$P_4^\pm$& $\frac{3\pi}{4}$& $\pm \cos
^{-1}\left(\frac{m}{n}\right)$ &$-\pi<\pm \cos
^{-1}\left(\frac{m}{n}\right)\leq \pi, n\neq 0$\\[0.2cm]
$P_5$& $\theta_1^\star$ & $0$
&$0\leq \theta_1^\star\leq \pi$\\[0.2cm]
$P_6$& $\theta_1^\star$ & $\pi$ &$0\leq \theta_1^\star\leq \pi$
\\[0.4cm]
\hline \hline
\end{tabular}
\end{center}
\end{table*}

\begin{table*}[t]
\caption[crit]{Stability of the critical points at infinity. We
use the notation $\delta=m^2-n^2$ and $\lambda^\pm=n\cos
\theta_1^\star\pm m\sin\theta_1^\star.$}
\label{critatinfinityprop}
\begin{center}
\begin{tabular}{@{\hspace{4pt}}c@{\hspace{14pt}}c@{\hspace{18pt}}c@{\hspace{18pt}}c@{\hspace{2pt}}}
\hline
\hline\\[-0.3cm]
Name & $(\lambda_1,\lambda_2)$ &$\rho'$&Stability\\[0.1cm]
\hline\\[-0.2cm]
$P_1^\pm$&  $(-n,n)$ & $>0$ & saddle\\[0.2cm]
$P_2^\pm$ & $(-n,n)$ &
$>0$  & saddle \\[0.2cm]
$P_3^\pm$&$\left(\frac{\sqrt{2}\delta}{n},\frac{\delta}{\sqrt{2}n}\right)$&$\left\{\begin{array}{cc}
  >0, &  \delta<0 \\
  <0, & \delta>0 \\
\end{array}\right.$&source if $n<0, n<m<-n$\\[0.2cm]
\quad & \quad& \quad & saddle otherwise\\[0.2cm]
$P_4^\pm$&$\left(-\frac{\sqrt{2}\delta}{n},-\frac{\delta}{\sqrt{2}n}\right)$
&$\left\{\begin{array}{cc}
  >0, &  \delta<0 \\
  <0, & \delta>0 \\
\end{array}\right.$& source if $n>0, -n<m<n$\\[0.2cm]
\quad  & \quad& \quad & saddle otherwise\\[0.2cm]
$P_5$& $\left(0,\lambda^+\right)$& $\left\{\begin{array}{cc}
  <0, &  \frac{\pi}{4}<\theta_1^\star<\frac{3\pi}{4} \\
  >0, & \text{otherwise} \\
\end{array}\right.$&nonhyperbolic\\[0.2cm]
$P_6$& $\left(0,\lambda^-\right)$&$\left\{\begin{array}{cc}
  <0, &  \frac{\pi}{4}<\theta_1^\star<\frac{3\pi}{4} \\
  >0, & \text{otherwise} \\
\end{array}\right.$&nonhyperbolic
\\[0.4cm]
\hline \hline
\end{tabular}
\end{center}
\end{table*}

In this section we investigate the dynamics at infinity by using
the central Poincar\'e projection method \cite{Lynch}.

To obtain the critical points at infinity we introduce spherical
coordinates ($\rho$ is the inverse of
$r=\sqrt{x_\phi^2+x_\vphi^2+y^2},$ then, $\rho\rightarrow 0$ as
$r\rightarrow \infty$):

\begin{align}
& x_\phi=\frac{1}{\rho}\sin\theta_1\cos\theta_2,\\
& y=\frac{1}{\rho}\sin\theta_1\sin\theta_2,\\
& x_\varphi=\frac{1}{\rho}\cos\theta_1
\end{align} where $0\leq \theta_1\leq \pi$ and $-\pi<\theta_2\leq
\pi,$ and $0<\rho<\infty.$

Defining the time derivative $f'\equiv
\rho\mathrm{d}f/\mathrm{d}\tau,$ the system
(\ref{eqxphi}-\ref{eqy}), can be written as

\begin{widetext}

\be \rho'=\frac{1}{2} \left(\cos ^2\theta_1-\cos (2\theta_2) \sin
^2\theta_1\right)+2 n \cos\theta_1 \sin ^2\theta_1 \sin^2\theta_2
\rho +O\left(\rho ^2\right)\label{radial}. \ee

and

\begin{align}
&\theta_1'=n \cos (2\theta_1) \sin \theta_1 \sin ^2\theta_2-\cos
\theta_1 \sin\theta_1 \sin
   ^2\theta_2 \rho +O\left(\rho ^2\right),\nonumber\\
   &\theta_2'=(n \cos \theta_1 \cos \theta_2+
   m \sin \theta_1) \sin \theta_2-\cos \theta_2 \sin \theta_2 \rho
   +O\left(\rho ^2\right).\label{eqangular}
\end{align}

\end{widetext}

Since equation (\ref{radial}) does not depends of the radial
component at the limit $\rho\rightarrow 0,$ we can obtain the
critical points at infinity by solving equations (\ref{eqangular})
in the limit $\rho\rightarrow 0.$ Thus, the critical points at
infinite must satisfy the compatibility conditions

\begin{eqnarray} & \cos (2 \theta_1) \sin
   \theta_1 \sin ^2\theta_2=0,\nonumber\\
   & (n \cos \theta_1 \cos
   \theta_2+m \sin \theta_1) \sin \theta_2=0.
   \label{compatibility}
\end{eqnarray}

First, we examine the stability of the pairs
$(\theta_1^\star,\theta_2^\star)$ satisfying the compatibility
conditions (\ref{compatibility}) in the plane
$\theta_1$-$\theta_2$, and then, we examine the global stability
by substituting in (\ref{radial}) and analyzing the sign of
$\rho'(\theta_2^\star,\theta_2^\star).$ In table
\ref{critatinfinity} it is offered information about the location
and existence conditions of these critical points. In table
\ref{critatinfinityprop} we summarize the stability properties of
these critical points.

The cosmological solutions associated to the critical points
$P_1^\pm$ and $P_2^\pm$ have the evolution rates
$\dot\phi^2/V=0,\, \dot\phi/\dot\varphi=0$ and
$H/\dot\varphi\equiv\rho/\sqrt{6}\rightarrow 0.$ \footnote{Do not
confuse $\rho$ with the matter energy density, the latter denoted
by $\rho_{\rm M}.$} These solutions are always saddle points at
infinity. The critical points $P_3^\pm$ and $P_4^\pm$ are sources
provided $n<0, n<m<-n$ or $n>0, -n<m<n,$ respectively. They are
saddle points otherwise. The associated cosmological solutions to
$P_3^\pm$ have the evolution rates
$\dot\phi^2/V=\frac{2m^2}{n^2-m^2},\, \dot\phi/\dot\varphi=-m/n,$
and $H/\dot\phi\equiv -n\rho/(\sqrt{3} m)\rightarrow 0,$ and
$H/\dot\varphi\equiv \rho/\sqrt{3}\rightarrow 0,$ whereas the
associated cosmological solutions to $P_4^\pm$ have the evolution
rates $\dot\phi^2/V=\frac{2m^2}{n^2-m^2},\,
\dot\phi/\dot\varphi=-m/n,$ and $H/\dot\phi\equiv n\rho/(\sqrt{3}
m)\rightarrow 0,$ and $H/\dot\varphi\equiv
-\rho/\sqrt{3}\rightarrow 0.$ The curves of critical points $P_5$
and $P_6$ are nonhyperbolic. The associated cosmological solutions
have expansion rates (valid for $\theta_1^\star\neq\pi/4$)
$V/\dot\phi^2=0, \dot\phi/\dot\varphi=\tan \theta_1^\star,
H/\dot\varphi=\rho \sec\theta_1^\star/\sqrt{6}\rightarrow 0,$ and
$V/\dot\phi^2=0, \dot\phi/\dot\varphi=-\tan \theta_1^\star,
H/\dot\varphi=\rho \sec\theta_1^\star/\sqrt{6}\rightarrow 0,$
respectively.

Concluding, although there exists unbounded orbits towards the
past, by examining the orbits at infinity, we get that the sources
satisfy the evolution rates
$\dot\phi^2/V\sim\frac{2m^2}{n^2-m^2},\, \dot\phi/\dot\varphi\sim
-m/n,$ with $H/\dot\phi\rightarrow 0,$ provided $n<0,
n<m<-n$ or $n>0, -n<m<n.$

\section{Concluding Remarks}\label{conclusions}

In this paper we have investigated the past asymptotic dynamics of
quintom cosmologies (which is closely related with the behavior of
typical orbits near an hyperbolae of critical points) by using the
Normal Form Theorem as a tool. We  have developed the general
(arbitrary order) normal expansion for the quintom model. This
expansion is the minimal in the sense that the terms that are
involved are the ``essential'' degrees of nonlinearity. The other
higher order terms are removed by making properly coordinate
transformation. By integrating the resulting system one is able to
construct unstable and center manifolds. We have explored the structure of unstable and center manifold of the origin up to
fourth order. From the structure of the unstable manifold we see
that that the past asymptotic behavior of quintom cosmologies,
with exponential potentials, is given  by a massless scalar field
cosmology for an open set of orbits. We provide here an
approximated formula for the center manifold. By examining the
center manifold we find that, up to fourth order, the curve of
critical points $C_-$ is not the past asymptotic global attractor. 

To get more accuracy we need to deal with the expansion up to arbitrary order $N>3.$ However, it is very difficult to get the expansion coefficients for $N>3.$ Theoretically, if the solution of (\ref{approxN1}) for $y_1$ admits a prolongation to $-\infty<\tau<\infty,$ (for instance, if $y_{10}=0$), then the trajectory passing through $(y_{10},y_{20},y_{30})$ at $t_0$ approaches the origin as $\tau\rightarrow -\infty,$ provided $\lambda_3>0.$ Otherwise, if the maximal interval, $(\alpha,\beta),$ has $\alpha$ finite, then, the orbits diverges in a finite time has $\tau\rightarrow \alpha^+.$ This fact is supported by the numerical simulations in reference \cite{Lazkoz:2006pa} (it appears that an open set of orbits escape to infinity into the past, which means that the local sources could not be the past attractor).

By examining the dynamics at infinity, using the central Poincar\'e projection method, we get that the sources satisfy the evolution rates $\dot\phi^2/V\sim\frac{2m^2}{n^2-m^2},\, \dot\phi/\dot\varphi\sim -m/n,$ with $H$ an infinitesimal of $\dot\phi,$ provided $n<0, n<m<-n$ or $n>0, -n<m<n.$
These results complete the past asymptotic analysis of quintom cosmologies.

\vskip .1in \noindent {\bf {Acknowledgments}}

The authors have the financial support of the MES of Cuba.

\appendix

\section{Normal Forms for vector fields}\label{NFTheory}

In this section we offer the main techniques for the construction
of normal forms for vector fields in $\mathbb{R}^n$. We follow the
approach in \cite{arrowsmith}.

Let ${\bf X}:\mathbb{R}^n\rightarrow \mathbb{R}^n$ be a smooth
vector field satisfying ${\bf X}({\bf 0})={\bf 0}.$ We can
formally construct the Taylor expansion of ${\bf x}$ about ${\bf
0},$ namely, ${\bf X}={\bf X}_1+{\bf X}_2+\ldots +{\bf
X}_k+{O}(|{\bf x}|^{k+1}),$ where ${\bf X}_r\in H^r,$ the real
vector space of vector fields whose components are homogeneous
polynomials of degree $r$. For $r=1$ to $k$ we write

\ben
&{\bf X}_r({\bf x})=\sum_{m_1=1}^{r}\ldots\sum_{m_n=1}^{r}\sum_{j=1}^{n}{\bf X}_{{\bf m},j}{{\bf x}}^{{\bf m}}{\bf e}_j,\nonumber\\
& \sum_i m_i=r, \een

Observe that ${\bf X}_1={\bf DX(\mathbf{0})}{\bf x}\equiv {\bf
A}{\bf x},$ i.e., the matrix of derivatives.

The aim of the normal form calculation is to construct a sequence
of transformations which successively remove the non-linear term
${\bf X}_r,$ starting from $r=2.$

The transformation themselves are of the form

\be {\bf x}={\bf y}+{\bf h}_r ({\bf y}),\label{htransform}\ee
where ${\bf h}_r\in H^r,\,r\geq 2.$

The effect of (\ref{htransform}) in ${\bf X}_1$ is as follows
\cite{arrowsmith}: Observe that ${\bf x}={O}(|{\bf y}|).$ Then,
the inverse of (\ref{htransform}) takes the form \be {\bf y}={\bf
x}-{\bf h}_r({\bf x})+{O}(|{\bf
x}|^{r+1}).\label{inversehtransform}\ee

By applying total derivatives in both sides, and assuming ${\bf
x}'={\bf A}{\bf x}+{\bf X}_r({\bf x})$, we find

\be {\bf y}'={\bf A} {\bf y}-{\bf L_A} {\bf h}_r ({\bf y})+{\bf
X}_r({\bf y})+{O}(|{\bf y}|^{r+1})\label{evoly}\ee where ${\bf
L_A}$ is the linear operator that assigns to ${\bf h(y)}\in H^r$
the Lie bracket of the vector fields ${\bf A y}$ and ${\bf h(y)}$:

\ben {\bf L_A}: H^r& &\rightarrow H^r\nonumber\\
     {\bf h}  & & \rightarrow  {\bf L_A} {\bf h (y)}={\bf D h(y)} {\bf A y}- {\bf A h(y)}.
\een

Both ${\bf L_A}$ and ${\bf X}_r\in H^r,$ so that the deviation of
the right-hand side of (\ref{evoly}) from ${\bf A y}$ has no terms
of order less than $r$ in $|{\bf y}|.$ This means that if ${\bf
X}$ is such that ${\bf X}_2=\ldots {\bf X}_{r-1}=0,$ they will
remain zero under the transformation (\ref{htransform}). This
makes clear how we may be able to remove ${\bf X}_r$ from a
suitable choice of ${\bf X}_r.$

The proposition 2.3.2 in \cite{arrowsmith} states that if the
inverse of ${\bf L_A}$ exists, the differential equation \be {\bf
x}'={\bf A}{\bf x}+{\bf X}_r({\bf x})+{O}(|{\bf
x}|^{r+1})\label{ODE1}\ee with ${\bf X}_r\in H^r,$ it is
transformed to \be {\bf y}'={\bf A y}+{O}(|{\bf
y}|^{r+1})\label{ODE2}\ee by the transformation (\ref{htransform})
where \be {\bf h}_r({\bf y})={\bf L_A}^{-1} {\bf X}_r(\bf
y)\label{Transform2}\ee

The equation \be {\bf L_A}{\bf h}_r({\bf y})={\bf X}_r(\bf y)\ee
is named the homological equation.

If ${\bf A}$ has distinct eigenvalues $\lambda_i,\, i=1,2,3,$ its
eigenvectors form a basis of $\mathbb{R}^n.$ Relative to this
eigenbasis, ${\bf A}$ is diagonal. It can be proved (see proof in
\cite{arrowsmith}) that ${\bf L_A}$ has eigenvalues $\Lambda_{{\bf
m},i}={\bf m}\cdot {\mathbf{\lambda}}-\lambda_i=\sum_j
m_j\lambda_j-\lambda_i$ with associated eigenvectors ${\bf
x^m}{\bf e}_i.$ The operator, ${\bf L_A}^{-1},$ exists if and only
if the $\Lambda_{{\bf m},i}\neq 0,$ for every allowed ${\bf m}$
and $i=1\ldots r.$

If we were able to remove all the nonlinear terms in this way,
then the vector field can be reduced to its linear part
$${\bf x}'={\bf X}({\bf x})\rightarrow {\bf y}'={\bf A}{\bf y}.$$
Unfortunately, not all the higher order terms vanishes by applying
this transformations. Particularly if resonance occurs.

The n-tuple of eigenvalues ${\bf
\lambda}=(\lambda_1,\ldots,\lambda_n)^T$ is resonant of order $r$
(see definition 2.3.1 in \cite{arrowsmith}) if there exist some
${\bf m}=(m_1,m_2,\ldots m_n)^T$ (a n-tuple of non-negative
integers) with $m_1+m_2+\ldots m_n=r$ and some $i=1\ldots n$ such
that $\lambda_i={\bf m}\cdot \lambda,$ i.e., if $\Lambda_{{\bf
m},i}=0$ for some ${\bf m}$ and some $i.$

If there is no resonant eigenvalues, and provided they are
different, we can use the eigenvectors of ${\bf A}$ as a basis for
$H^r.$ Then, we can write ${\bf h}_r$ as
$${\bf h}_r({\bf x})=\sum_{{\bf m},i,\sum m_j=r}h_{{\bf m},i}{\bf x}^{\bf m}{\bf e}_i$$ and any vector
field ${\bf X}\in H^r$ as $${\bf X}({\bf x})=\sum_{{\bf m},i,\;
\sum m_j=r} {\bf X}_{{\bf m},i}{\bf x}^m {\bf e}_i$$ where  ${\bf
m}=(m_1,m_2,\ldots,m_n)^T,$ ${\bf
x}^m=x_1^{m_1}\,x_2^{m_2}\,\ldots\,x_n^{m_n}$ and ${\bf
e}_i,\,i=1,\ldots n$ stands for the canonical basis in
$\mathbb{R}^n.$ If the eigenvalues of ${\bf A}$ are not resonant
of order $r,$ then $$h_{{\bf m},i}=X_{{\bf m},i}/\Lambda_{{\bf
m},i}.$$ This gives ${\bf h}_r$ explicitly in terms of ${\bf
X}_r.$

In the case of resonance occurs, we proceed as follows. If ${\bf
A}$ can diagonalized, then the eigenvectors of ${\bf L}_{\bf A}$
form a basis of $H^r.$ The subset of eigenvectors of ${\bf L}_{\bf
A}$ with non-zero eigenvalues then form a basis of the image,
$B^r,$ of $H^r$ under ${\bf L}_{\bf A}.$ It follows that the
component of ${\bf X}_r$ in $B^r$ can be expanded in terms of
these eigenvectors and ${\bf h}_r$ chosen such that $$h_{{\bf
m},i}=X_{{\bf m},i}/\Lambda_{{\bf m},i}.$$ to ensure the removal
of these terms. The component, ${\bf w}_r,$ of ${\bf X}_r$ lying
in the complementary subspace, $G^r,$ of $B^r$ in $H^r$ will be
unchanged by the transformations ${\bf x}={\bf y}+{\bf h}_r({\bf
y})$ obtained from $B^r.$

Since $$ {\bf X}_r({\bf y}+{\bf h}_{r+k}({\bf y}))={\bf X}_r({\bf
y})+{O}(|{\bf y}|^{r+k+1}),r\geq 2, k=1,2,\ldots,$$ these terms
are not changed by subsequent transformations to remove
non-resonant terms of higher order.

The above facts are expressed in

\begin{thm}[theorem 2.3.1 in \cite{arrowsmith}]\label{NFTheorem}
Given a smooth vector field $\bf X({\bf x})$ on $\mathbb{R}^n$
with ${\bf X(0)=0},$ there is a polynomial transformation to new
coordinates, ${\bf y},$ such that the differential equation ${\bf
x}'={\bf X}({\bf x})$ takes the form ${\bf y}'={\bf J}{\bf
y}+\sum_{r=1}^N {\bf w}_r({\bf y})+{O}(|{\bf y}|^{N+1}),$ where
${\bf J}$ is the real Jordan form of ${\bf A}={\bf D X}({\bf 0})$
and ${\bf w}_r\in G^r,$ a complementary subspace of $H^r$ on
$B^r={\bf L_A}(H^r).$
\end{thm}

\end{document}